\documentclass[article]{elsarticle}
\usepackage{mathtools}
\usepackage{amssymb}
\usepackage{lineno,hyperref}
\usepackage{graphicx}
\usepackage{subcaption}
\usepackage{float}
\modulolinenumbers[5]
\journal{European Polymer Journal}

\begin{document}

\begin{frontmatter}

\title{Effect of Chain Length Distribution on Mechanical Behavior of Polymeric Networks}

\author{Mohammad Tehrani}

\author{Alireza Sarvestani\corref{cor1}}
\ead{sarvesta@ohio.edu}
\cortext[cor1]{Corresponding author}

\address{Department of Mechanical Engineering, Ohio University, Athens OH 45701, USA}

\begin{abstract}
The effect of network chain distribution on mechanical behavior of elastomers is one of the long standing problems in rubber mechanics. The classical theory of rubber elasticity is built upon the assumption of entropic elasticity of networks whose constitutive strands are of uniform length. The kinetic theories for vulcanization, computer simulations, and indirect experimental measurements all indicate that the microstructure of vulcanizates is made of polymer strands with a random distribution of length. The polydispersity in strand length is expected to control the mechanical strength of rubber as the overloaded short strands break at small deformations and transfer the load to the longer strands. The purpose of this contribution is to present a simple theory of rubber mechanics which takes into account the length distribution of strands and its effect on the onset of bulk failure.
\end{abstract}

\begin{keyword}
\texttt rubber elasticity, chain length distribution, softening, polydispersity
\end{keyword}

\end{frontmatter}

\section{Introduction}
\noindent A major assumption of classical rubber elasticity is the monodispersity of the constitutive strands, the sub-chains between the two consecutive crosslink points. The conventional crosslinking techniques, however, are essentially uncontrolled processes and hence, the formation of ideal monodisperse networks is not probable. Direct measurement of randomness in internal structure of rubber compounds is unfeasible due to insolubility of the polymer networks. First efforts to indirectly quantify the structural polydispersity of vulcanizates go back to the pioneering works of Toboslky \cite{tobolsky}, Beuche \cite{bueche}, Gehman \cite{gehmanmolecular}, and Watson \cite{berry}, using relaxometry of stressed networks or measurement of swelling pressure.

\bigskip
\noindent The effect of  strand polydispersity on the overall mechanical behavior of polymer networks is of great research and technological importance. The simplest polydisperse networks can be formed by end-linking of functionally terminated crosslinkers with a multimodal length distribution. Mark and his co-workers conducted a comprehensive study on bimodal polymer networks (\cite{mark} and the references therein). Their results point to a great enhancement in ultimate mechanical properties of the network, namely an increase in the toughness and larger elongation at break. These findings were attributed to the distribution of stress between the short and long chains. The enhancement in strength is primarily due to the limited deformability of non-Gaussian short chains. Following the rupture of short chains, the stress is transferred to the long strands which exhibit larger deformation at break.

\bigskip
\noindent In vulcanizates, the strand length distribution is expected to be non-uniform and range from very short to very long strands \cite{gehmannetwork}. This assumption is validated by a number of computer simulation studies. Grest and Kremer \cite{grestkremer}, for example, simulated the equilibrium structure of randomly crosslinked networks with the number of crosslinks  well above the percolation threshold. The network was formed by instantaneous crosslinking of long primary chains in a melt state. In the ideal case of completely random crosslinking, the association of chains can be regarded as statistically independent events. Theoretically, this means that the distribution of crosslink points along the primary chains must be Gaussian and hence, the distribution of strand length between crosslink points must follow a simple exponential form with a decay length. The simulation results of Grest and Kremer support these predictions.

\bigskip
\noindent Modeling fracture and mechanical failure of polymer networks continues to be a subject of ongoing research \cite{volokh2010modeling,trapper2008cracks,dal2009micro,miehe2014phase}. The ultimate mechanical properties of polymer networks are affected by a host of influences, ranging in a wide spectrum of length scales. This includes the microstructure of a single polymer chain (e.g., helicity, interatomic potentials, crosslinking, isomerization, etc.) as well as the chain's local environment (entanglements, cracks, etc.). The focus of current study is to provide a theoretical basis to evaluate the role of strand polydispersity in the bulk failure of polymer networks.\footnotemark 
\footnotetext{Following Volokh \cite{volokh2010modeling}, here, the concept of bulk failure refers to the \textquotedblleft continuum damage mechanics\textquotedblright in which the material failure is controlled by damage accumulation and evolution of internal structure of the bulk material. This approach is different from the so called \textquotedblleft cohesive zone\textquotedblright models in which the properties of bulk material remain unchanged and fracture is presented by introducing interface cohesive elements whose behavior is controlled by some traction-seperation laws.}
The importance of strand length distribution for the mechanical strength of vulcanizates was first highlighted by Gehman \cite{gehmannetwork}. He proposed that upon deformation of a random network, shorter strands break at considerably smaller deformations compared to the longer ones. This deformation-induced network alteration continues concurrent with increasing deformation and controls the onset of mechanical failure. This proposition is adopted here and forms the basis of a micromechanical model for the elasticity and damage initiation in elastomers with a random distribution of strand length. This study is inspired by the recent work of Itskov and Knyazeva \cite{itskovrubber} who proposed a model for rubber elasticity based on the chain length statistics. Here, their approach is advanced by introducing a failure criterion based on the interatomic pair potential and considering damage accumulation using a simple first-order kinetic theory.

\section{Model}
\noindent Bueche \cite{bueche} and Watson \cite{watson1953chain,watson1954chain} originally proposed an expression for the strand length distribution function in a random network. Consider a network formed by vulcanization of infinitely long polymer chains. Let ${n_{j}}$ be the number of strands with ${j}$ statistical segments. If the placement of crosslink points is taken to be completely random with probability ${p}$, then the probability distribution of having a strand with ${j}$ statistical segments, $P(j)$, can be expressed as
 
\begin{equation} 
P(j)=\frac{n_{j}}{\sum_{j}^{}n_{j}}=(1-p)^{j-1} p
\end{equation}

\noindent where ${\sum_{j}^{}n_{j}}$ is the total number of existing strands. The assumption of completely random placement of crosslinks warrants the probability ${p}$ to be a constant and equal to the reciprocal of average strand length $\overline{j}=\frac{1}{p}$. At the limit of large $\overline{j}$ values, Eq. (1) leads to a distribution function
 
\begin{equation} \label{eq:LeghGe}
P(j)=\frac{1}{m}\Big(1+\frac{1}{m}\Big)^{-j}
\end{equation}

\noindent where $m=\frac{1}{p}-1$. For large $m$ values, the approximation $\Big(1+\frac{1}{m}\Big)^{m}\approx e$ holds and thus the distribution (2) accept a simple exponential form

\begin{equation} 
P(j)=\frac{1}{\overline{j}}e^{-j/\overline{j}}
\end{equation}

\noindent Note that Eq. (3) represents the probability distribution of strand length in an ideally random crosslinked network, where the position of crosslinks are taken to be statistically independent.\par

\bigskip
\noindent Now consider a network of crosslinked flexible strands subjected to a quasi-static finite deformation. To keep the formulation simple, throughout this paper it is assumed that the network is incompressible, although extension
of the presented theory to the compressible networks is
possible. The end-to-end vector of each strand in the reference and current configurations is represented by $\textbf{R}_{0}$ and $\textbf{R}$, respectively (Figure 1). The network is formed by random crosslinking of the strands whose length follow distribution (3). The effect of other structural properties such as crystallinity or entanglement are not taken into account. The conformational entropy of a strand with $j$ statistical segments, stretched by $\lambda$, is

\begin{equation} 
S(\lambda,j)=-jk_{B}\bigg (\frac{\lambda}{\sqrt{j}}\beta+\ln\frac{\beta}{\sinh\beta}\bigg )-S_{0}
\end{equation}

\noindent where 

\begin{equation} 
\beta=\pounds^{-1}\bigg (\frac{\lambda}{\sqrt{j}}\bigg ) \quad
\end{equation}

\noindent Here $k_{B}$ is the Boltzmann constant, $S_{0}$ is the reference conformational entropy, and $\pounds^{-1}$ stands for the inverse Langevin function. This way, the free energy of each strand can be written as

\begin{equation} 
w(\lambda,j)=U-TS(\lambda,j)
\end{equation}

\noindent where $T$ is the absolute temperature and $U$ is the internal energy controlled by the interatomic interactions. Following a classical approach in rubber elasticity, here the contribution of internal energy in the free energy landscape is ignored. The rupture of strands, however, is essentially controlled by the nature of this interatomic potential, as described later. Thus \cite{arruda1993three}

\begin{equation} 
w(\lambda,j)=jk_{B}T\bigg (\frac{\lambda}{\sqrt{j}}\beta+\ln\frac{\beta}{\sinh\beta} \bigg)+w_{0}
\end{equation}

\noindent where $w_{0}$ represents the deformation-independent part of the free energy.

\bigskip
\noindent To obtain the free energy density function for a network of strands, the so called chain orientation distribution function $C(\theta,\phi)$ is used. It represents the probability distribution of having a strand with end-to-end vector $\textbf{R}$ at spherical coordinates $\theta$ and $\phi$ in the current configuration. Hence, \cite{wu1993improved}

\begin{equation} 
\int\limits_{0}^{\pi}\int\limits_{0}^{2\pi}   C(\theta,\phi) \ \sin\theta \ d\theta \ d\phi=1 
\end{equation}

\noindent The free energy density function of an ensemble of deformed strands with polydisperse length, occupying volume $\textit{V}$, can be obtained as

\begin{equation} 
W(\lambda)=\frac{\sum n_{j}}{\textit{V}}\int\limits_{0}^{2\pi}\int\limits_{0}^{\pi}\int\limits_{1}^{\infty} \ P(j) \ w \big (\lambda(\theta,\phi),j \big ) \ C(\theta,\phi) \  \sin\theta \ dj \ d\theta \ d\phi
\end{equation}

\noindent A similar function, $C_{0}(\theta_{0},\phi_{0})$,  can be defined for the orientation of strands at the reference configuration. Assuming that the strands orientation is initially random, this probability distribution can be characterized by $C_{0}(\theta_{0},\phi_{0})=\frac{1}{4\pi}$. It thus follows that \cite{wu1993improved}

\begin{equation}
C(\theta,\phi)=C_{0}\frac{sin\theta_{0}}{sin\theta}J^{-1}
\end{equation}

\noindent where $J$ is the Jacobian of deformation gradient. After substitution of (10) into (9) and taking advantage of incompressibility condition, one can obtain

\begin{equation} 
W(\lambda)=\frac{\sum n_{j}}{4\pi\textit{V}}\int\limits_{0}^{2\pi}\int\limits_{0}^{\pi}\int\limits_{1}^{\infty} \ P(j) \  w(\lambda,j) \ \sin\theta_{0} \ dj \ d\theta_{0} \ d\phi_{0}
\end{equation}

\noindent The stretch along an arbitrary direction can be expressed in the reference configuration and in terms of the macroscopic principal stretches, $\lambda_{i}$, as

\begin{equation}\lambda^{2} (\theta_{0},\phi_{0})=(\lambda_{1}  \sin\theta_{0}  \cos\phi_{0})^{2}+(\lambda_{2}  \sin\theta_{0}   \sin\phi_{0})^{2}+(\lambda_{3}  \cos\theta_{0})^{2}
\end{equation}

\bigskip
\noindent Eq. (11) represents the elastic energy of a network with polydisperse strands where all strands are assumed to be elastically active. The free energy function presented by Eq. (7) accounts for the finite extensibility of flexible strands and diverges as the stretch approaches the ultimate locking value of $\lambda_{lock}=\sqrt{j}$ \cite{arruda1993three}. Assuming that crosslink points move in an affine fashion, shorter strands are expected to experience a larger entropic tension. As proposed by Itskov and Knyazeva \cite{itskovrubber}, the highly extended strands snap at some finite stretch and become elastically inactive. Therefore, at each direction, strands shorter than a certain length break and do not contribute to the energy function (11). The ultimate strength of an elastically active strand is determined either by scission of bonds along the backbone or cleavage of a crosslink. The activation energy for rupture is directly related to the nature of interatomic potential or the dissociation energy of crosslink coagents. Different harmonic and anharmonic potential functions have been used to present the energy landscape of interatoimic dissociation in polymer chains \cite{doerr1994breaking,oliveira1994breaking,crist1984polymer,dal2009micro}. The Morse potential, for example, is used to predict the stiffness of a covalent bond in a nan-Gaussian polymer chain during cleavage \cite{garnier2000covalent}. Here, a Morse pair-potential is used to describe the energy barrier for debonding

\begin{equation} \label{eq:LeghGe}
U(r)=U_{0}\Big(1-e^{-\alpha(r-r_{0})}\Big)^2
\end{equation}

\noindent where $U_{0}$ is the dissociation energy, $\alpha$ is a constant that determines bonds elasticity, and $r$ and $r_{0}$ show the deformed and undeformed (equilibrium) length of a bond, respectively (Figure 2). The strand rupture occurs when the applied force exceeds the critical value of

\begin{equation} \label{eq:LeghGe}
\Big(f_{M}\Big)_{max}=\frac{\alpha U_{0}}{2}
\end{equation}

\noindent beyond which the bonds become unstable. Eq. (14) limits the maximum force that can be developed in each strand. It is assumed that the bond cleavage occurs when this force equals the restoring entropic force between the crosslinks, $f_{e}$, defined as

\begin{equation}
f_{e}=\frac{\partial w}{\partial R}
\end{equation}

\noindent Using (14) and (15), one can find the length of the shortest strand,  $j_{min}$, that withstands the macroscopic stretch $\lambda$ without rupturing. That is

\begin{equation} \label{eq:LeghGe}
j_{min}(\lambda)=\frac{\lambda^{2}(\theta_{0},\phi_{0})}{\xi}
\end{equation}

\noindent with

\begin{equation} 
\frac{1}{\xi}=\frac{3(3+\sqrt{4\gamma+9})}{2\gamma^{2}}+1 \qquad , \quad \gamma=\frac{\alpha a  U_{0}}{2k_{B}T}
\end{equation}

\noindent where $a$ is the characteristic length of one statistical segment. Since only elastically active strands contribute to stress production, the lower limit of the first integral in Eq. (11) can be replaced with $j_{min}$

\begin{equation} 
W(\lambda)=\mu\int\limits_{0}^{2\pi}\int\limits_{0}^{\pi}\int\limits_{j_{min}(\lambda)}^{\infty} \ P(j) \ w(\lambda,j) \ \sin\theta_{0} \ dj \ d\theta_{0} \  d\phi_{0}
\end{equation}

\noindent where $\mu=\frac{\sum n_{j}}{4\pi\textit{V}}$. Using the spectral decomposition theorem, the respective Cauchy stresses of the incompressible network can be derived from the strain energy density function $W(\lambda)$ as \cite{ogden1997non}

\begin{equation} 
\boldsymbol{\sigma}= \sum_{k=1}^{3} \lambda_{k} \frac{\partial W}{\partial\lambda_{k}} (\boldsymbol{n}^{(k)}\otimes \boldsymbol{n}^{(k)})  
\end{equation} \label{eq:DefStress}

\noindent where $\lambda_{k}$ and $\textbf{n}^{(k)}$ are the eigenvalues and eigenvectors of the right stretch tensor, respectively. Substitution of Eq. (18) into (19) yields

\begin{equation} \label{eq:LeghBC2}  
\boldsymbol{\sigma}= \mu\sum_{k=1}^{3} \lambda_{k}(\boldsymbol{n}^{(k)}\otimes \boldsymbol{n}^{(k)})  \  \Big(\int\limits_{0}^{2\pi}\int\limits_{0}^{\pi}  \frac{\partial}{\partial \lambda_{k}}  \int\limits_{j_{min}(\lambda)}^{\infty} \ P(j) \ w(\lambda,j)  \ \sin\theta_{0} \ dj \ d\theta_{0} \ d\phi_{0}\Big)
\end{equation}

\noindent from which all components of the Cauchy stress tensor can be evaluated (see the Appendix).

\bigskip
\noindent The proposed formulation can be readily generalized to include the effect of history-dependent damage in a random network subjected to a cyclic loading. From the standpoint of thermal fluctuation theory, the history-dependent damage in solids is controlled by the elementary events of bond rupture and the failure is ensued by damage accumulation in the solid. Therefore, it is assumed that the number of elastically active strands with $j$ statistical segments is a function of time, presented by $n_{j,t}$. The kinetics of irreversible bond rupture can be represented by a first-order kinetic process proposed by Eyring \cite{krausz1975deformation}

\begin{equation} 
\frac{dn_{j,t}}{dt}=-k_{r}n_{j,t} 
\end{equation}

\noindent where $k_{r}$ shows the frequency of bond rupture in elastically active strands. Using the well-known Zhurkov formula \cite{zhurkov1966kinetic}

\begin{equation} 
k_{r}=k_{r0} \ \textrm{exp}\big[f_{e}\delta/k_{B}T\big ]
\end{equation}

\noindent Here, $k_{r0}$ is a rate constant and $\delta$ is an activation length. Substitution of Eq. (22) into (21) and solving for  $n_{j,t}$ yield

\begin{equation} 
\theta_{j}(\lambda,j,t)=\frac{n_{j,t}}{n_{j,0}}=\textrm{exp}\big[\varXi(\lambda,t]\big ]
\end{equation} 

\noindent with

\begin{equation} 
\varXi(\lambda,t)=-\int\limits_{0}^{\overline{t}} \ \textrm{exp} \big[\beta(\lambda(\overline{t}))\overline{\delta} \ \big] d\overline{t}
\end{equation}

\noindent where $\overline{t}=t \ k_{r0}$ and $\overline{\delta}=\delta / a$. Here $n_{j,0}$ shows the number of elastically active strands with $j$ statistical segments before loading. Assuming

\begin{equation}
P(j)=\frac{n_{j,0}}{\sum_{j}^{} n_{j,0}}
\end{equation} 

\noindent now the time-dependent strain energy density function can be written as

\begin{equation} 
W(\lambda,t)=\mu\int\limits_{0}^{2\pi}\int\limits_{0}^{\pi}\int\limits_{j_{min}}^{\infty} \ \theta_{j}(\lambda,j,t) \ P(j)  \ w(\lambda,j) \sin\theta_{0}  \ dj \ d\theta_{0} \ d\phi_{0}
\end{equation}

\noindent with $\mu=\frac{\sum_{j}^{} n_{j,0}}{4\pi\textit{V}}$.

\section{Results}
\noindent This part presents some examples of model predictions for the elasticity and strength of polymer networks with random structure, represented by the chain length distribution (3). This numerical study aims to reveal how two major model parameters control the network behavior: the average length index ($\overline{j}$) and the bond strength parameter ($\xi$). Figure 3, shows the effect of $\overline{j}$ on the stress response of random networks subjected to quasi-static uniaxial tension and simple shear deformations. The numerical calculation of the integrals appearing in Eqs. (20) and (24) is carried out by MATLAB. The assumed values for $\overline{j}$ are chosen to be comparable with the simulation results of Svaneborg  et al. \cite{svaneborg2005disorder}. Polydisperse networks with smaller $\overline{j}$ values include a larger population of short chains. Despite their slightly higher stiffness at small to moderate stretches, these networks show lower ultimate strength compared to those with larger average strand length. These conclusions can be explained considering the finite extensibility and non-Gaussian behavior of shorter strands manifested at small stretches. With increasing the applied deformation, the shorter strands gradually approach their contour length and ultimately fail under the high entropic tension. The progressive degradation of network leads to material softening and controls the ultimate strength of the network.

\bigskip
\noindent The structure-properties of random networks have been the subjects of a number of molecular simulation studies \cite{grestkremer,svaneborg2005disorder,gavrilov2014computer}. The results reflect both the microstructural details and the macroscopic stress developed in the networks. Interestingly, there is a reasonable agreement between the results of  simple probability distribution (3) and the distribution of strand lengths in the idealized simulations \cite{grestkremer,gavrilov2014computer}. Figure 4 shows the stress-stretch curves in so called Mooney-Rivlin coordinates predicted by the proposed model in comparison with the simulation results of Gavrilov and Chertovich \cite{gavrilov2014computer} for random networks. They used dissipative particle dynamics to simulate the structure of randomly crosslinked polymer chains, including the effects of Langevin statistics and finite extensibility of strands. The results show an initial hardening stage due to the non-Gaussian response of the short strands, with a good agreement up to the macroscopic stretch of $\lambda\approx2.5$. The rupture of elastically active strands is a feature of the present model that is not considered in Gavrilov-Chertovich simulation. As a result, the present model predicts a drastic softening due to progressive rupture of strands whereas the predicted stress in their simulation remains practically unbounded.\par

\bigskip
\noindent Figure 5 compares the model predictions for the ultimate principal stretches (corresponding to the maximum Cauchy stress) in various plane stress loading modes  with the experimental data of Hamdi et al. \cite{hamdi2006fracture} on SBR. Seeking a generalized failure criterion at multiaxial quasi-static loadings, Hamdi et al. \cite{hamdi2006fracture}  used defect-free vulcanizates and measured the elongation at break of samples subjected to uniaxial or biaxial deformations. The biaxial tests were conducted on membrane-like samples by inflation of thin membranes in elliptical meniscuses with different aspect ratios to obtain different biaxiality ratios. It appears that both $\overline{j}$ and $\xi$ are able to significantly push the failure envelope and a wide range of experimental data could be covered by changing the value of these parameters. The strength parameter $\xi$ reflects the dissociation energy between the monomers or the crosslinking coagents. Figure 6 shows that the network strength strongly depends on $\xi$ and a small change in dissociation energy leads to a significant alteration in mechanical strength. The relative importance of networks randomness and binding energy (represented by $\overline{j}$ and $\xi$, respectively) in determination of network strength depends on the microstructural details and the nature of chemical reactions used to form the network. For example, it is well-known that carboxylated, sulfuric, and carbon-to-carbon crosslinked vulcanizates show markedly different strengths under tension \cite{bateman1963chemistry,kok1986effects}. Despite the higher dissociation energy of direct carbon-carbon bonds, however, the peroxide cures generally exhibit lower mechanical strength compared to the rubber vulcanized by accelerated sulfur \cite{gehman1969network,kok1986effects}. This strength inferiority is rooted in the significant randomness in the internal structure, introduced by peroxide reaction. Dicumyl peroxide is a vulcanizing agent that exclusively reacts with polyisoprene by abstraction of $\alpha$-methylenic hydrogen atoms. As shown by Park and Lorenz \cite{parks1963effect}, the decomposed peroxides form isoprene radicals that contribute in crosslinking with a very high efficiency. With increasing the probability of crosslinking, the population of short chains increases at the expense of strength, in accordance with the results of the presented model.\par

\bigskip
\noindent Finally, the model is used to predict history-dependent damage and stress-induced degradation of polydisperse networks during a cyclic loading. Figure 7 presents the stress-stretch behavior of two random networks with average chain length of 20 and 100. The networks are subjected to a constant amplitude periodic stretch, as shown in Figure 7(a). The results feature the well-known and frequently reported characteristics of rubber hysteresis \cite{ayoub2011modeling,ayoub2014visco}. The most apparent is the gradually decreasing global stiffness of the network concurrent with increasing the loading cycles. The dissipation at the first few cycles is significantly larger than the energy loss associated with the following cycles. Indeed, the hysteresis practically disappears after just a few number of deformation cycles with a constant amplitude. The energy dissipation and degradation of network mechanical properties are more pronounced in networks with smaller $\overline{j}$. These results collectively suggest that randomness in internal structure and polydispersity in strands length contribute to the fatigue behavior and could effectively limit the average lifetime of the networks.

\section{Concluding Remarks}

\noindent In an attempt to correlate the ultimate macroscopic mechanical properties of polymeric networks to their internal structure, a theory of rubber elasticity is formulated in which the network microstructure is random and the strands are polydisperse in length. On the basis of a simple statistical analysis, new expressions for the strain energy density function and the Cauchy stress tensor are obtained that take into account the strand length distribution and predict its effect on the bulk damage in the network. Strands with different lengths respond differently to the applied macroscopic deformation. Short chains quickly experience the Langevin effect and break under a relatively small stretch. The progressive failure of the shorter strands continues and eventually determines the ultimate strength of the network. Direct mechanical measurements are insufficient to exclusively provide any information on the strand length distribution and the randomness in internal structure of polymer networks. Thus, the value of presented model is that it can be used to test the validity of assumptions made about the network statistics at the microscale by comparing the model predictions with the relevant experimental data.

\bigskip
\noindent Certain remarks must be made with regard to the validity and capability of the proposed approach. First, the validity of simple statistical model presented by Eq. (3) depends on a major assumption that all statistical segments have an equal chance to contribute to crosslinking reactions. While this assumption may be acceptable for peroxide cures, it does not do justice to the complicated structure of sulfur vulcanizates. Vulcanization of natural rubber with accelerated sulfur is essentially an autocatalytic reaction \cite{gehman1969network,alfrey1963kinetics,arends1963general}. Sulfur facilitates the local reactions adjacent to a crosslink and leads to increased functionality and further enhancement of the network strength. Second, the assumption that all strands in a polydisperse network follow an affine deformation is not backed by a rigorous justification. Very short strands, say with just a few statistical segments in length, hardly act as elastically active chains \cite{tobolsky,bueche} and thus their failure under small deformation is unlikely. This becomes particularly important when considering the distribution function (3) in which network strands of short size could be in abundance. Third, while the energetic Morse potential is used to predict the bond rupture, the contribution of enthalpic interactions in the free energy (Eq. (6)) is disregarded. It is known that the consideration of enthalpic contributions removes the singularity caused by the Langevin effect at large deformations. Inclusion of enthalpic contributions in calculation of free energy is expected to provide a more realistic estimation of the network strength.

\section{Appendix}

\noindent Eq. (20) can be written as

\begin{multline} \tag{A1}
\boldsymbol{\sigma}=\mu \sum_{k=1}^{3} \lambda_{k}(\boldsymbol{n}^{(k)}\otimes  \boldsymbol{n}^{(k)})  \   \Big(\int\limits_{0}^{2\pi}\int\limits_{0}^{\pi} \int\limits_{j_{min}(\lambda)}^{\infty} \ P(j) \  \frac{\partial}{\partial \lambda_{k}}  w(\lambda , j) \ \sin\theta_{0} dj d\theta_{0} d\phi_{0}  \\
- \int\limits_{0}^{2\pi}\int\limits_{0}^{\pi}  \frac{\partial j_{min}(\lambda)}{\partial \lambda_{k}} \ P(j_{min}) \   w(\lambda,j_{min}) \ \sin\theta_{0} dj d\theta_{0} d\phi_{0}  \Big)
\end{multline}

\noindent where

\begin{equation} \tag{A2}
\frac{\partial w(\lambda_{r},j)}{\lambda_{k}}=\frac{\partial \lambda}{\partial \lambda_{k}} \frac{\partial}{\partial \lambda}  w(\lambda)
\end{equation}

\noindent From Eq. (12), it follows that

\begin{equation} \tag{A3}
\frac{\partial \lambda}{\partial \lambda_{1}}=\frac{\lambda_{1}}{\lambda}\sin^{2}\theta_{0} \cos^{2}\phi_{0}
\end{equation}

\begin{equation} \tag{A4}
\frac{\partial \lambda}{\partial \lambda_{2}}=\frac{\lambda_{2}}{\lambda} \sin^{2}\theta_{0} \sin^{2}\phi_{0}
\end{equation}

\begin{equation} \tag{A5}
\frac{\partial \lambda}{\partial \lambda_{3}}=\frac{\lambda_{3}}{\lambda} \cos^{2}\theta_{0}
\end{equation}

\noindent Taking partial derivative of the strain energy of a single strand with respect to the macroscopic stretch yields 

\begin{equation} \tag{A6}
\frac{\partial}{\partial \lambda}  w(\lambda) =  \beta+\lambda\frac{\partial \beta}{\partial \lambda} +\frac{1}{\beta}\frac{\partial \beta}{\partial \lambda}-\frac{\partial \beta}{\partial \lambda}\coth(\beta)
\end{equation}

\noindent The inverse Langevin function can be approximated in different ways \cite{jedynak2015approximation}. Here, we used the so called Puso's approximation \cite{puso1994mechanistic}

\begin{equation} \tag{A7}
\beta=\pounds^{-1}(y)\approx\frac{3y}{1-y^{3}}
\end{equation}

\noindent which leads to

\begin{equation} \tag{A8}
\frac{\partial}{\partial y}\pounds^{-1}(y)\approx\frac{3+6y^3}{(1-y^{3})^{2}}
\end{equation}

\section*{References}
\bibliography{mybibfile}
\newpage
\begin{figure}[H]
	
	\includegraphics[width=\linewidth]{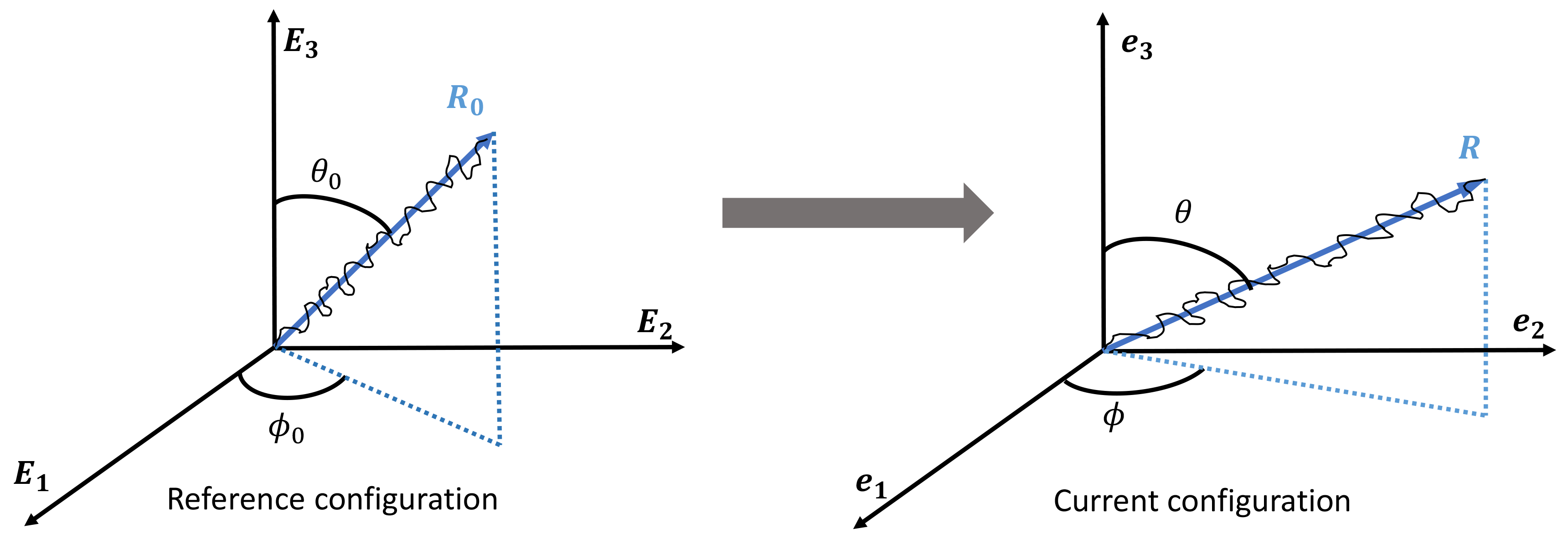}
	\caption{A network strand in the reference and current configurations.}
	\label{fig:configurations}
\end{figure}
\newpage
\begin{figure}
	\centering
	\includegraphics[width=0.75\linewidth]{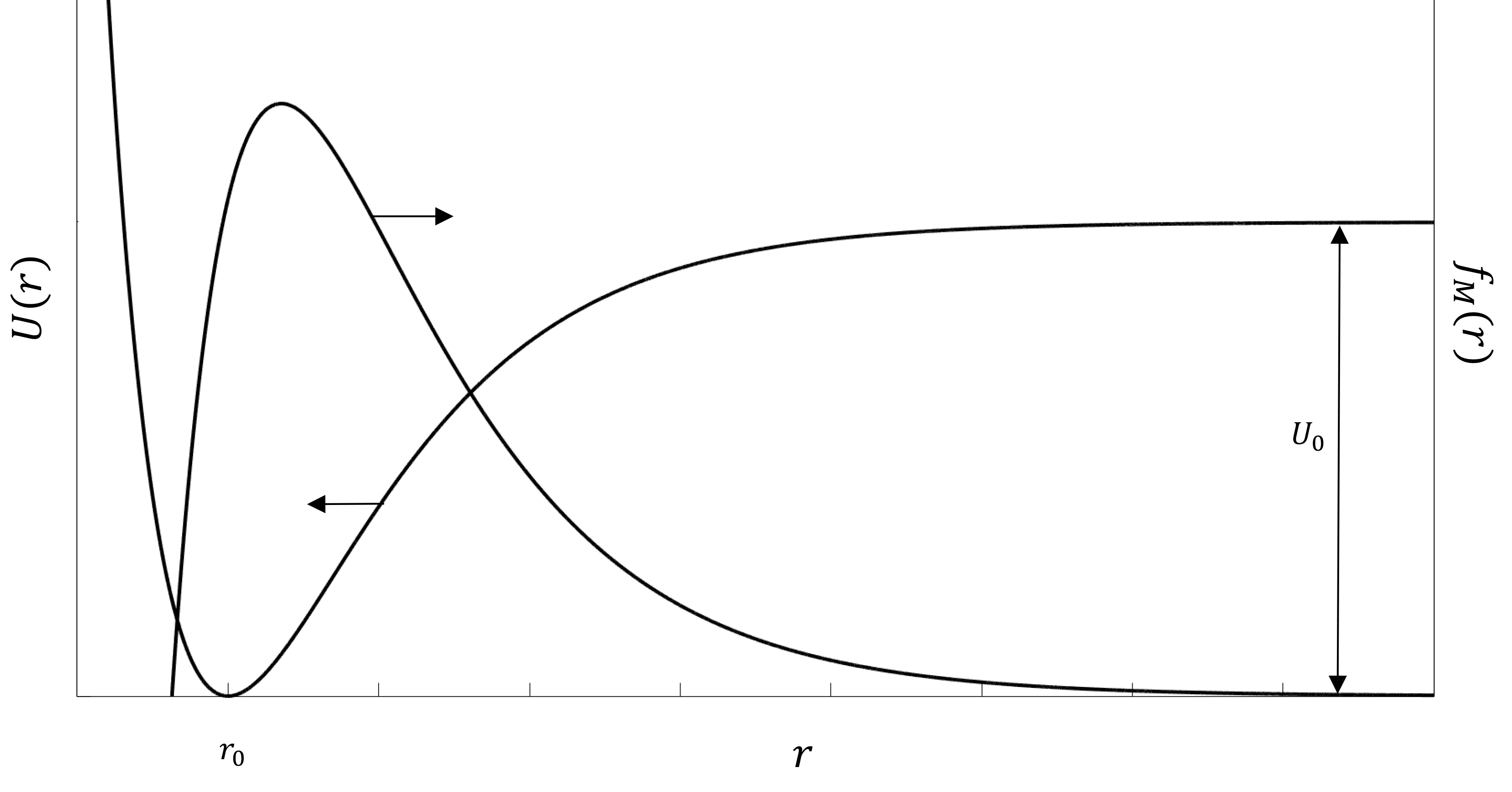}
	\caption{A Morse-type pair potential and the corresponding interatomic force.}
	\label{fig:morse}
\end{figure}
\newpage
\begin{figure}
	\centering
	\begin{subfigure}{1\textwidth}
		\centering
		\includegraphics[width=\linewidth]{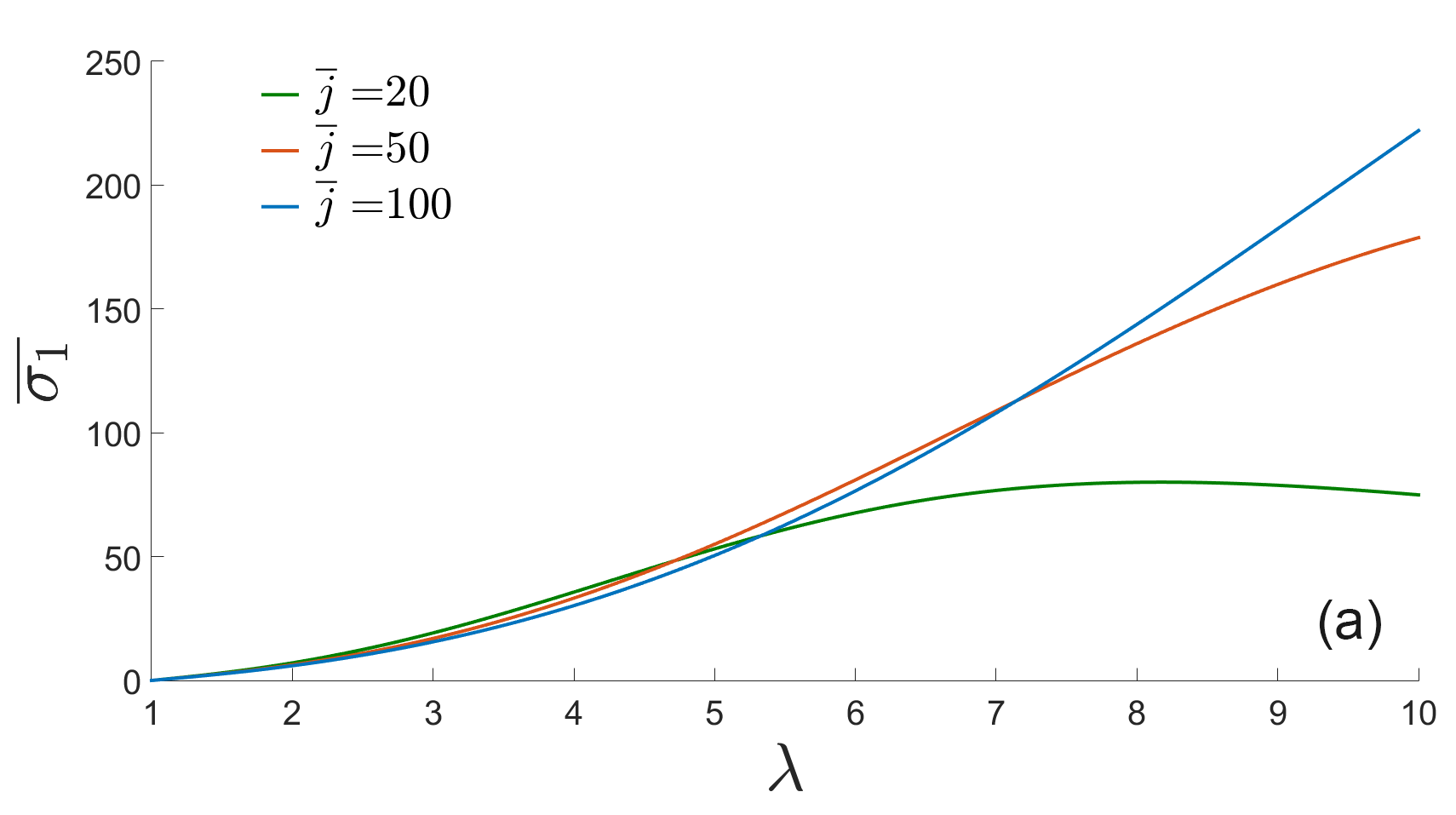}
	\end{subfigure}
	\begin{subfigure}{1\textwidth}
		\centering
		\includegraphics[width=\linewidth]{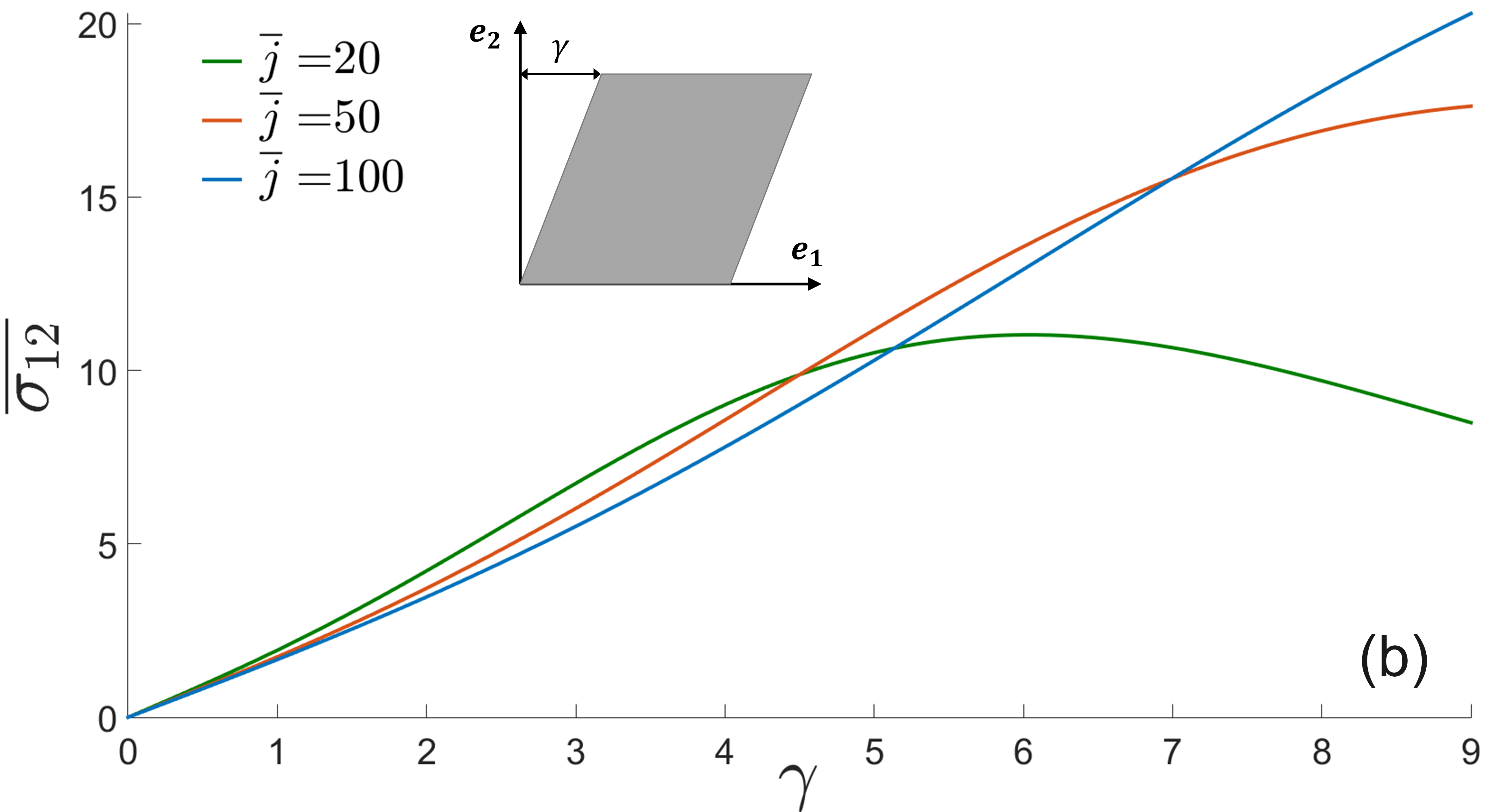}
	\end{subfigure}
	\caption{The effect of average strand length, $\overline{j}$, on the stress behavior of random networks. (a) Variation of normalized tensile stress with stretch in uniaxial tension ($\overline\sigma_{1}=\frac{\sigma_{1}-\sigma_{2}}{\mu k_{B}T}$ where $\sigma_{1}$ and $\sigma_{2}$ represent the principal stresses). (b) Variation of normalized shear stress with shear in simple shear deformation ($\overline\sigma_{12}=\frac{\sigma_{12}}{\mu k_{B}T}$). The bond strength parameter is taken to be $\xi=0.99$.}
\end{figure}
\newpage
\begin{figure}
	\includegraphics[width=\linewidth]{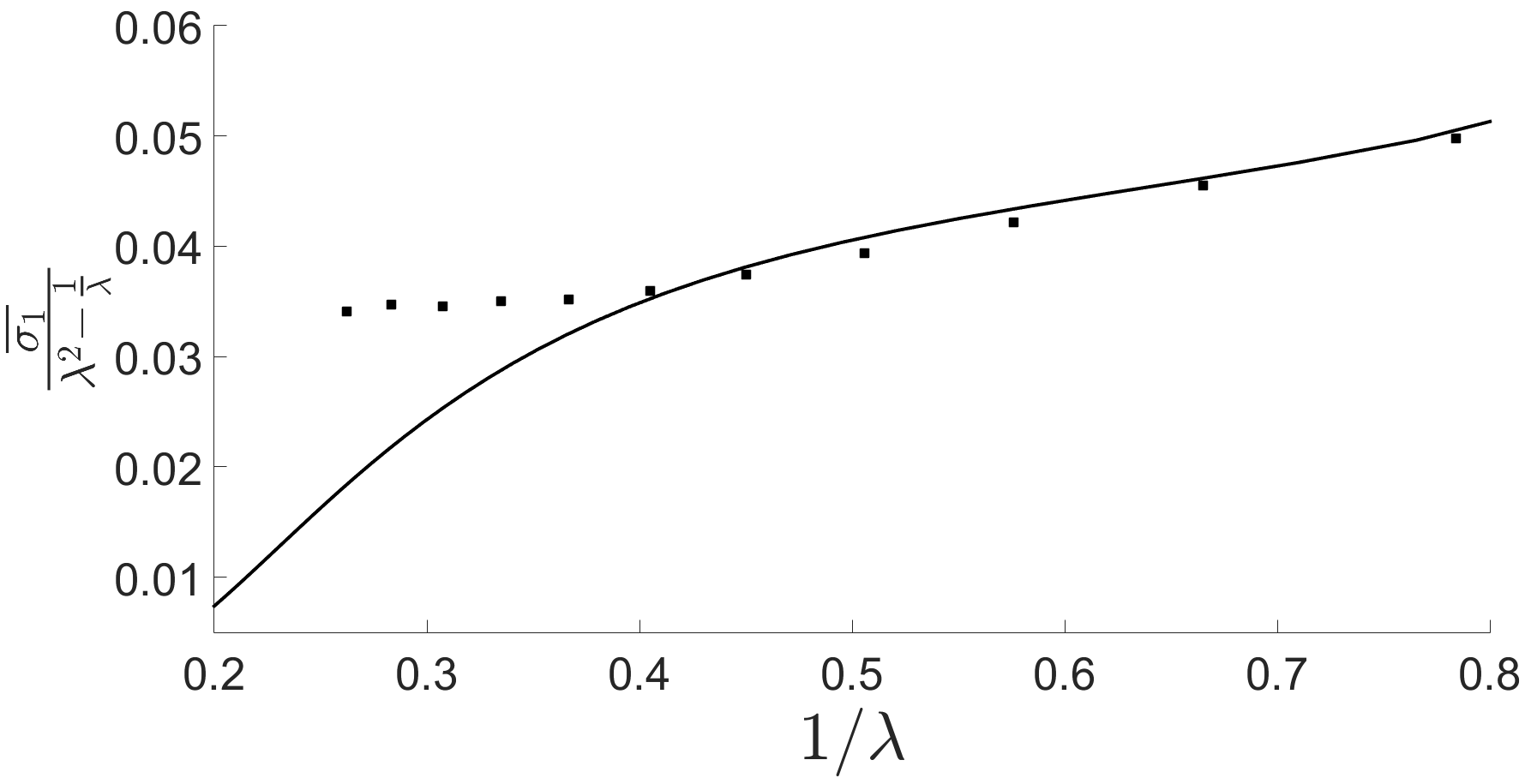}
	\caption{Comparison between the model prediction (solid line) and simulation results of Gavrilov-Chertovich \cite{gavrilov2014computer} for a random network  with $\overline{j}=10.54$ subjected to a uniaxial stress ($\overline\sigma_{1}=\frac{\sigma_{1}-\sigma_{2}}{\mu k_{B}T}$ where $\sigma_{1}$ and $\sigma_{2}$ represent the principal stresses). The bond strength parameter is taken to be $\xi=0.99$.}
	\label{fig:boat1}
\end{figure}
\newpage
\begin{figure}
	\centering
	\begin{subfigure}{0.75\textwidth}
		\centering
		\includegraphics[width=\linewidth]{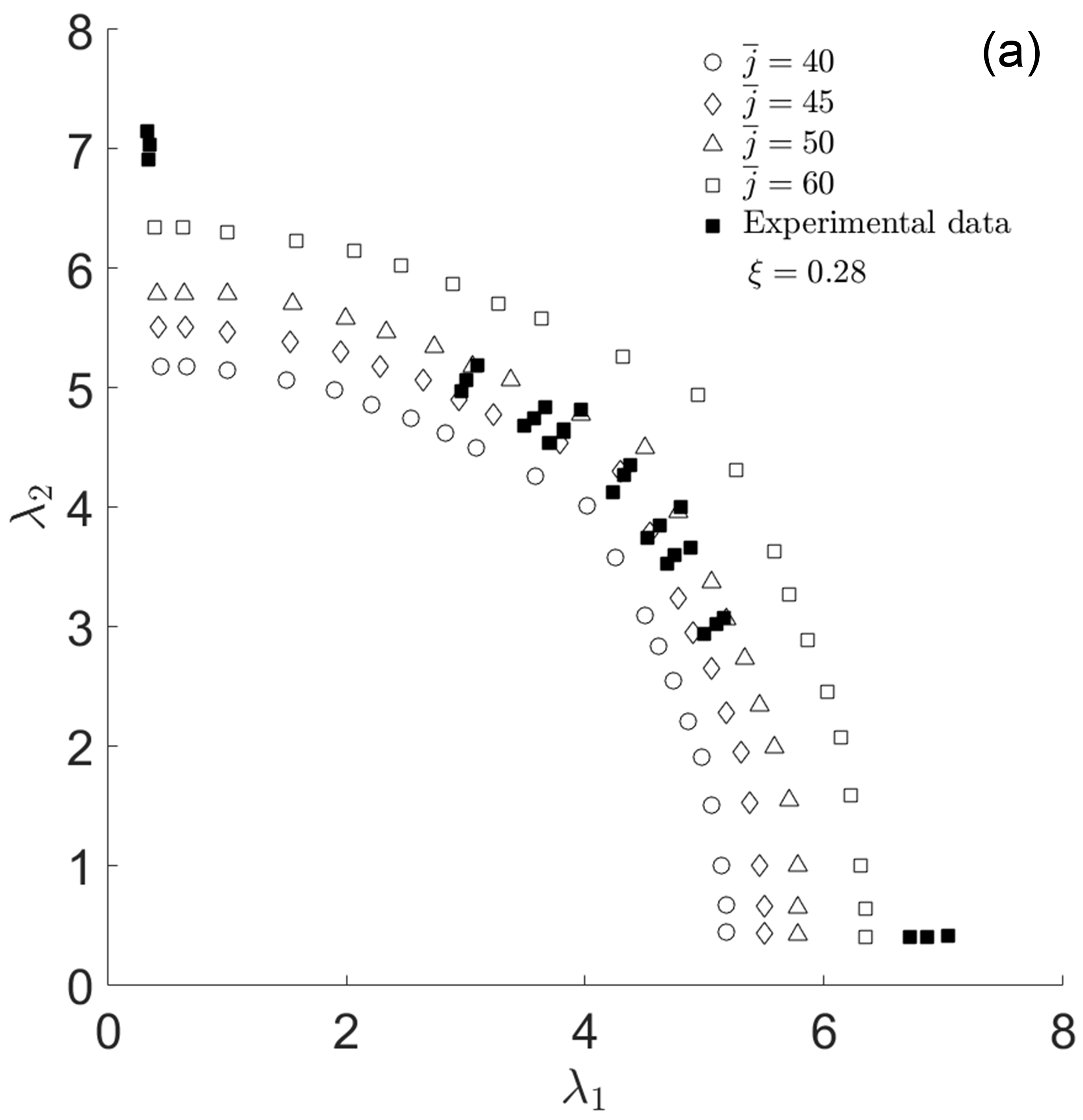}
	\end{subfigure}
	\begin{subfigure}{0.75\textwidth}
		\centering
		\includegraphics[width=\linewidth]{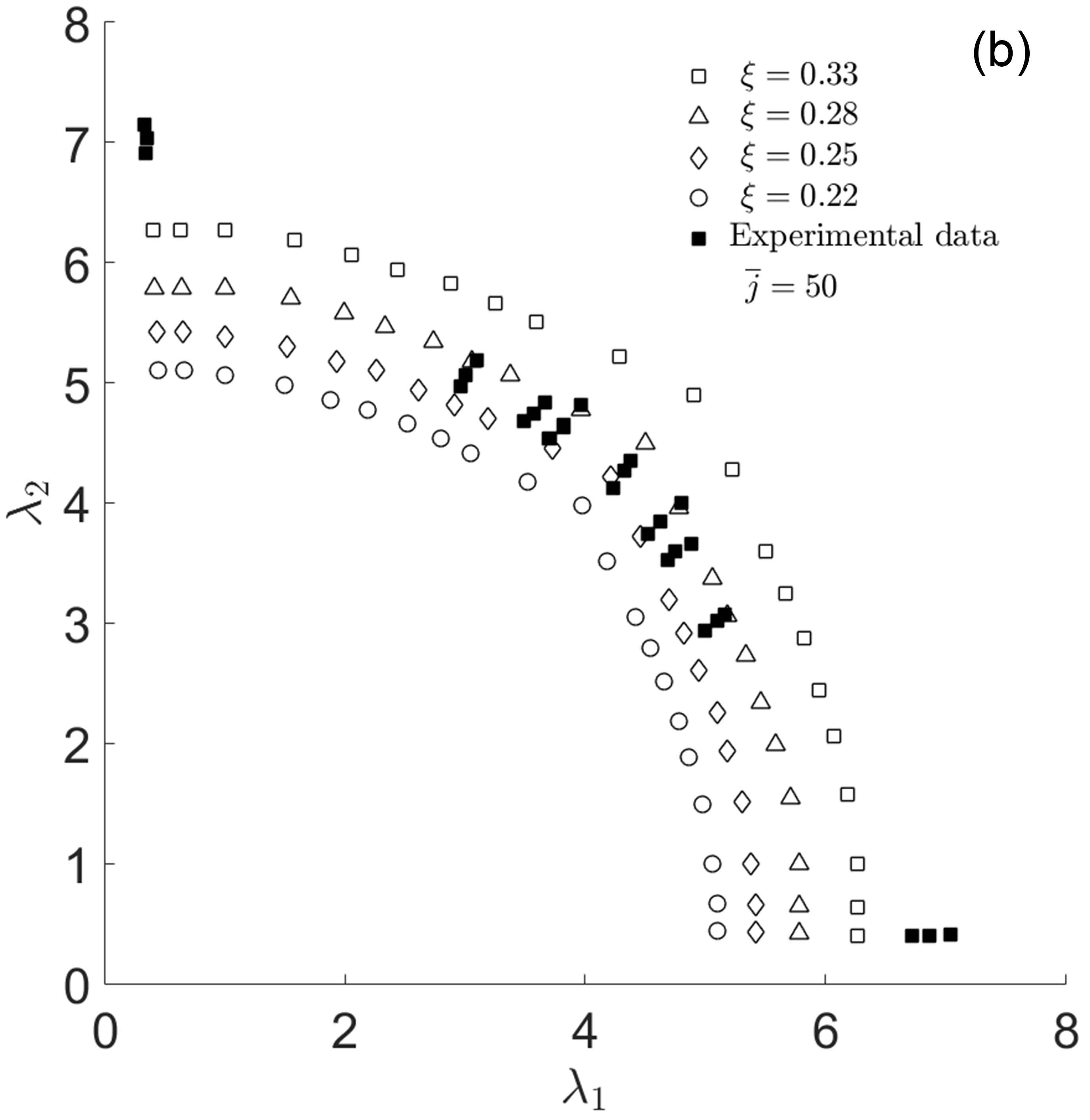}
	\end{subfigure}
	\caption{Effect of (a) average strand length, $\overline{j}$, and (b) bond strength parameter, $\xi$, on the ultimate stretches (corresponding to the maximum Cauchy stresses) as predicted by the proposed model. The results are compared with the experimental data of Hamdi et al. \cite{hamdi2006fracture} on SBR.}
\end{figure}
\newpage
\begin{figure}
	\centering
	\begin{subfigure}{1\textwidth}
		\centering
		\includegraphics[width=\linewidth]{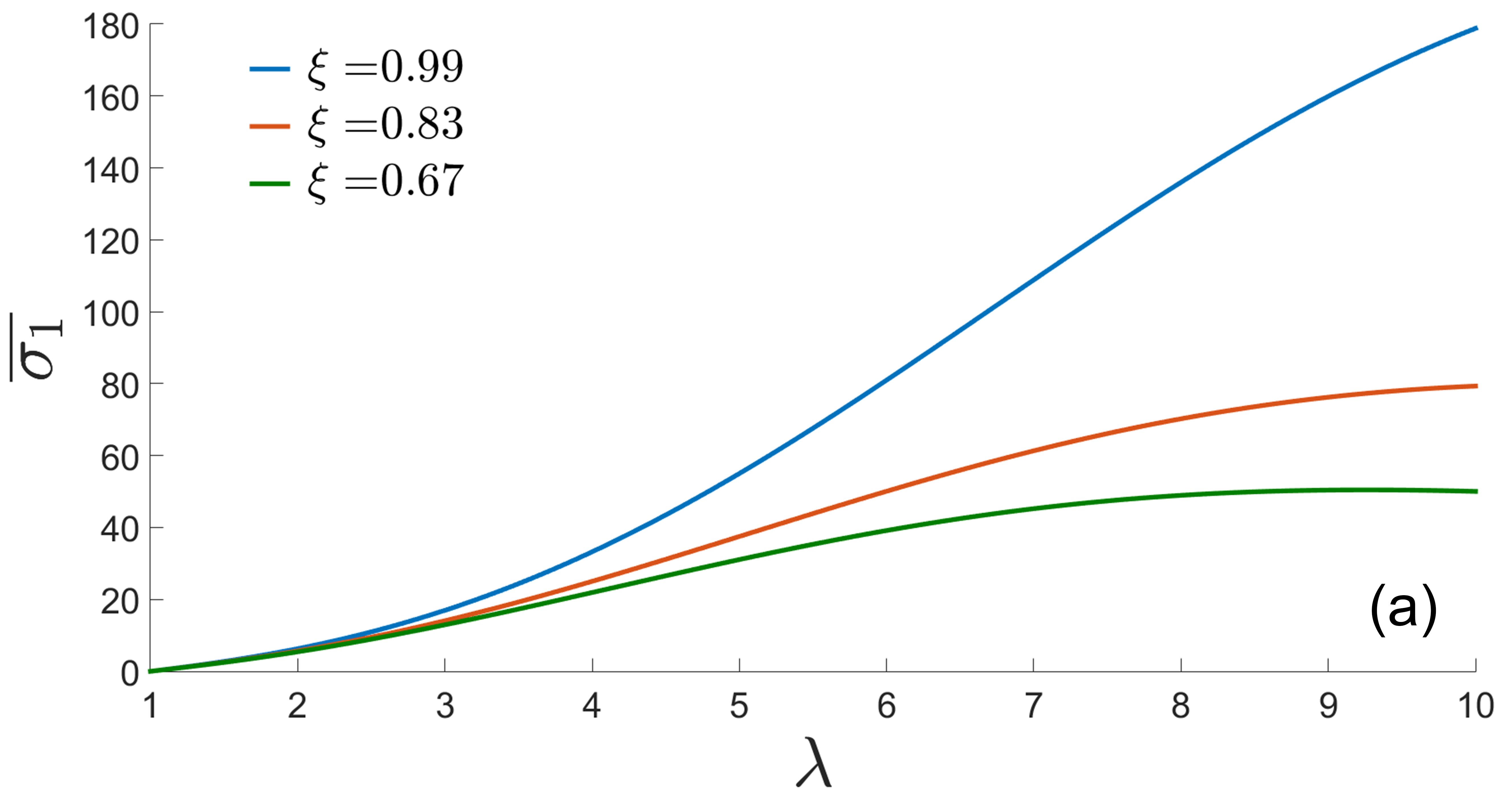}
	\end{subfigure}
	\begin{subfigure}{1\textwidth}
		\centering
		\includegraphics[width=\linewidth]{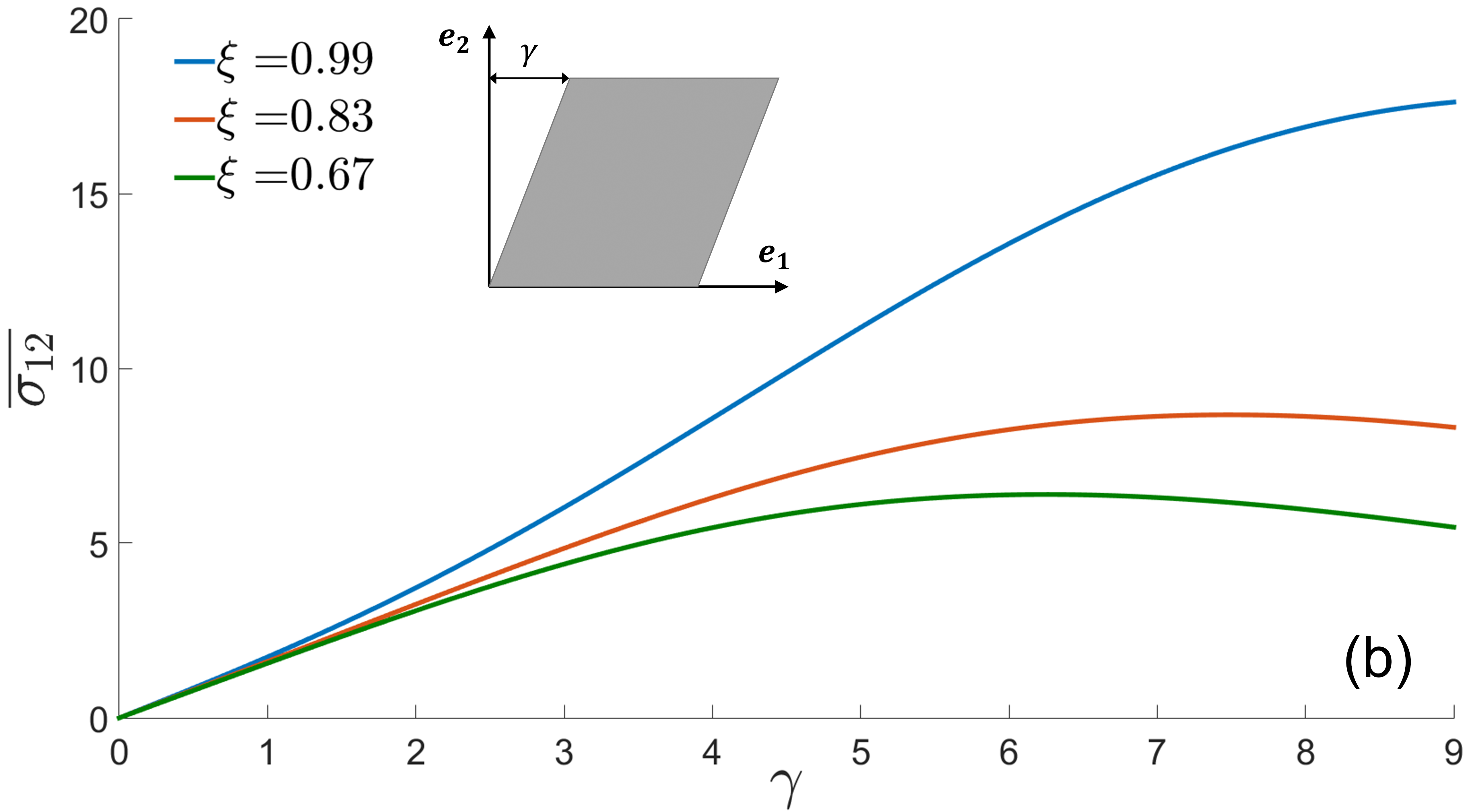}
	\end{subfigure}
	\caption{The effect of bond strength parameter, $\xi$, on the stress behavior of random networks. (a) Variation of normalized tensile stress with stretch in uniaxial tension ($\overline\sigma_{1}=\frac{\sigma_{1}-\sigma_{2}}{\mu k_{B}T}$ where $\sigma_{1}$ and $\sigma_{2}$ represent the principal stresses). (b) Variation of normalized shear stress with shear in a simple shear deformation ($\overline\sigma_{12}=\frac{\sigma_{12}}{\mu k_{B}T}$). The average strand length is taken to be $\overline{j}=20$.}
\end{figure}
\newpage
\begin{figure}
	\centering
	\begin{subfigure}{1\textwidth}
		\centering
		\includegraphics[width=\linewidth]{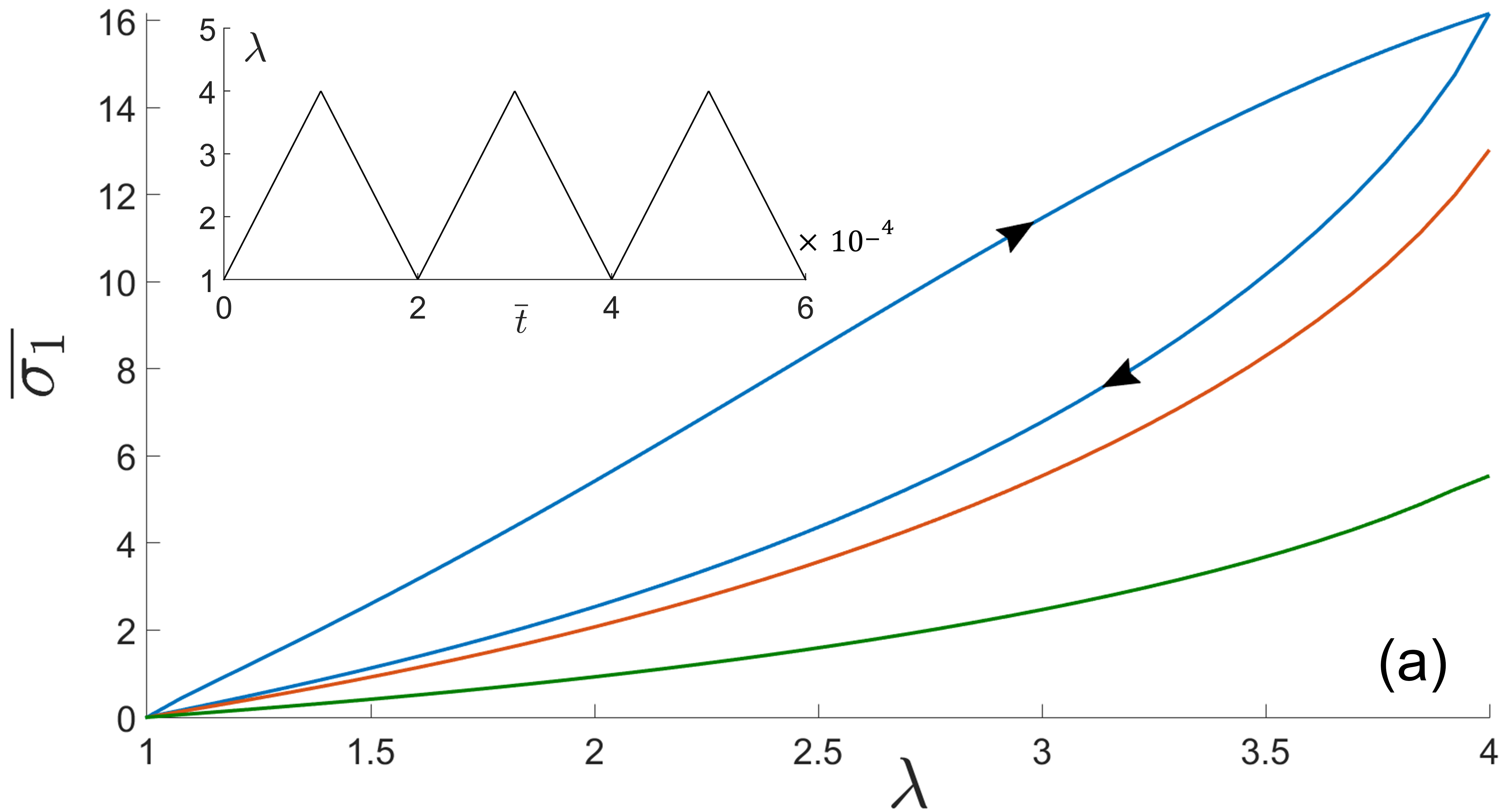}
	\end{subfigure}
	\begin{subfigure}{1\textwidth}
		\centering
		\includegraphics[width=\linewidth]{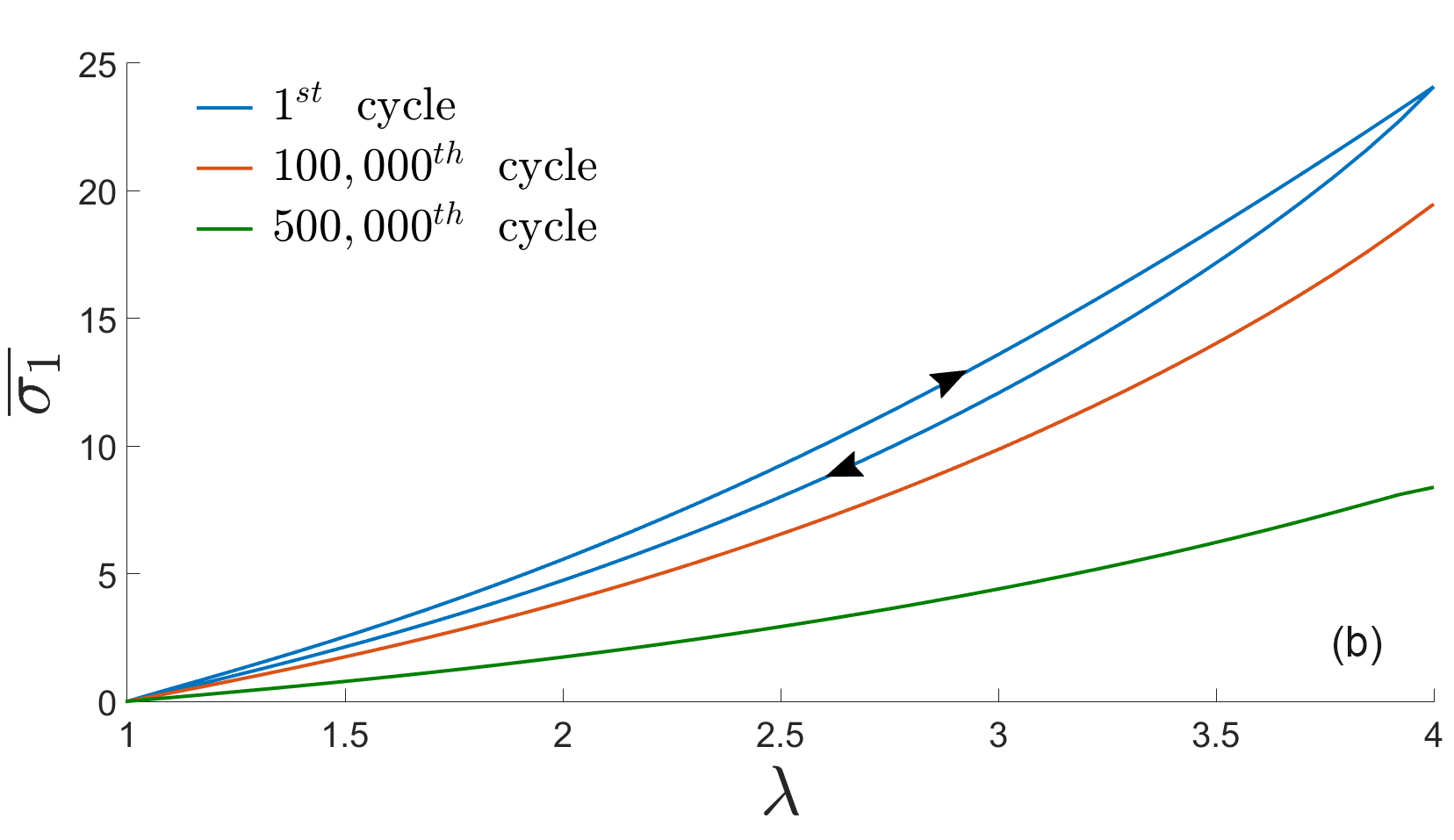}
	\end{subfigure}
	\caption{Variation of normalized stress with cyclic stretch in random networks with (a) $\overline{j}=20$ (b) and $\overline{j}=100$ ($\overline\sigma_{1}=\frac{\sigma_{1}-\sigma_{2}}{\mu k_{B}T}$ where $\sigma_{1}$ and $\sigma_{2}$ represent the principal stresses). The networks are subjected to slow cyclic axial stratching, as shown by inset. A full cycle of loading-unloading lasts 50 $\textrm{s}$. Other model parameters are $\xi=0.99$, $k_{r0}=2\times10^{-6} \ \textrm{s}^{-1}$ \cite{lavoie2}, and $\delta \approx a$ \cite{yarin1998model}.}
\end{figure}

\end{document}